\documentclass[aps,reprint,showpacs,superscriptaddress]{revtex4-1}
\usepackage{blindtext}
\usepackage{amssymb, amsmath,framed}

\usepackage{bm}
\usepackage[colorlinks=true,linkcolor=blue]{hyperref}
\expandafter\ifx\csname package@font\endcsname\relax\else
 \expandafter\expandafter
 \expandafter\usepackage
 \expandafter\expandafter
 \expandafter{\csname package@font\endcsname}
\fi
\def\bq{\begin{equation}}
\def\eq{\end{equation}}
\def\bqy{\begin{eqnarray}}
\def\eqy{\end{eqnarray}}

\def\p{\partial}

\def\rh{\rho}

\def\p{\partial}

\def\cala{\mathcal{A}}
\def\calb{\mathcal{B}}
\def\calc{\mathcal{C}}

\def\calk{\mathcal{K}}

\def\calm{\mathcal{M}}

\begin{document}
\title{Concomitant Hamiltonian and topological structures of extended magnetohydrodynamics}

\author{Manasvi Lingam}
\email{mlingam@princeton.edu}
\affiliation{Department of Astrophysical Sciences, Princeton University, Princeton, NJ 08544, USA}
\affiliation{Department of Physics and Institute for Fusion Studies, The University of Texas at Austin, Austin, TX 78712, USA}
\author{George Miloshevich}
\email{gmilosh@physics.utexas.edu}
\affiliation{Department of Physics and Institute for Fusion Studies, The University of Texas at Austin, Austin, TX 78712, USA}
\author{Philip J. Morrison}
\email{morrison@physics.utexas.edu}
\affiliation{Department of Physics and Institute for Fusion Studies, The University of Texas at Austin, Austin, TX 78712, USA}

\begin{abstract}
The paper describes the unique geometric properties of ideal magnetohydrodynamics (MHD), and demonstrates how such features are inherited by extended MHD, viz. models that incorporate two-fluid effects (the Hall term and electron inertia). The generalized helicities, and other geometric expressions for these models are presented in a topological context, emphasizing their universal facets. Some of the results presented include: the generalized Kelvin circulation theorems; the existence of two Lie-dragged 2-forms; and two concomitant helicities that can be studied via the Jones polynomial, which is widely utilized in Chern-Simons theory. The ensuing commonality is traced to the existence of an underlying Hamiltonian structure for all the extended MHD models, exemplified by the presence of a unique noncanonical Poisson bracket, and its associated energy. 
\end{abstract}

\maketitle

\section{Introduction} \label{SecIntro}

Ideal magnetohydrodynamics (MHD) is the simplest model in plasma physics, and is used extensively in the arenas of fusion, space and astrophysical plasmas; see e.g. \cite{GP04} and references therein. From a mathematical perspective, ideal MHD represents a natural extension of ideal hydrodynamics (HD) as it is endowed with geometric properties that mimic those of ideal HD. Much of this geometric structure arises from the flux freezing condition, which is intimately linked with the conservation of magnetic and cross helicities.

It is a widely accepted maxim that topological invariants play a key role in several areas of physics. In HD, the fluid helicity plays a similar role, as it constitutes a measure of the Gauss linking number of vortex lines, as shown in the pioneering work of \cite{M69}. In MHD, an equivalent role is played by the magnetic helicity, whose topological properties were extensively investigated in \cite{BF84}. Subsequently, the topological formulations of HD and MHD, especially their attendant helicities, underwent increasing mathematical sophistication; representative examples in this category include \cite{Moff90,FH91,MR92,MT92,Berg93,HoSc96,GHS97,AK98,RSB99,CDG01,BP06,TF06,Ghy07,Ric08,EPS12,WZ14}. Fluid/magnetic helicities also emerge naturally as a consequence of the underlying relabelling symmetry of HD and MHD, on account of Noether's theorem \cite{PM96,PAM96}. It is worth noting that the relativistic version of helicity \cite{Mah03}, and its concomitant topological properties have been studied in \cite{Mah03,YKY14,KYF14,Peg15}. 

We emphasize that this topological nature has come under greater experimental scrutiny \cite{KI13,SKPKI14,KSI14} demonstrating that helicity is converted from links/knots to coils. In addition to the importance of these helicities as topological invariants, they are also indispensable in understanding the self-organization and relaxation of fluids/plasmas \cite{W58,T74,FA85,T86,JPS95,Berg99,YHW10,Set15}. In the astrophysical context, (magnetic) helicity has played a pivotal role in solar physics \cite{Berg84,Low94,BAT09,Petal14}, helicity injection \cite{NZZ03,JC07}, reconnection \cite{Pont11,PWHG11}, turbulence \cite{Bisk03,BS05} and dynamo theory \cite{BS05,BSS12}. We observe that concurrent applications and uses of helicity in fusion plasmas also abound; some examples are listed in \cite{Boo86}.

Although MHD is endowed with several unique properties, it is also inapplicable in several domains. Hence, several extensions of ideal MHD have been studied, such as Hall MHD \cite{L60}, electron MHD \cite{GKR94}, and extended MHD \cite{L59}. There has been much interest in Hall MHD, as it possesses helicities and relaxed states akin to that of ideal MHD \cite{MY98,YM02}, and has been widely studied as a model of fast reconnection \cite{Bisk00}. Hall MHD can be further generalized to include the effects of electron inertia, thereby resulting in extended MHD. Alternatively, a model with electron inertia, but lacking the Hall terms, was proposed in \cite{KM14,LMT14} with the accompanying title of inertial MHD.

In this paper, we propose to highlight the commonality of all the extended MHD models through several avenues. These include the delineation of the appropriate conserved helicities and the appropriate frozen-in fluxes. Furthermore, we demonstrate that all of these models possess a virtually identical Hamiltonian structure \cite{AKY15,LMM15} -- the latter refers to the existence of a suitable (conserved) energy and a noncanonical Poisson bracket. Such Poisson brackets were first constructed for ideal HD and MHD in \cite{MG80}, and are quite different in structure as the physical Eulerian fields (such as density, velocity, etc.) are not canonical in nature. An extended discussion of these brackets, and the advantages of the Hamiltonian description of fluids/plasmas can be found in \cite{pjm82,HMRW85,M98}.

The outline of the paper is as follows. The common Hamiltonian structure of different extended MHD models is presented in Section \ref{SecHam}. In Section \ref{SecGTXMHD}, we sketch the unifying topological aspects of the various extended MHD models. Finally, we summarize our results in Section \ref{SecConc}, and indicate how they could play an important role in fusion and astrophysical plasmas.

\section{Hamiltonian structure of extended MHD models} \label{SecHam}
In this Section, we shall present the dynamical equations of different extended MHD models, demonstrate the existence of a common Hamiltonian structure, and thereby construct the associated helicities and generalized frozen-in fluxes.

\subsection{Mathematical preliminaries} \label{SSecMathP}
We begin with the equations of extended MHD, which comprise of the continuity equation
\begin{equation} \label{ContEq}
\frac{\p \rh}{\p t} + \nabla \cdot \left(\rh {\bf V}\right) = 0,
\end{equation}
the equation for the momentum density
\begin{eqnarray} \label{MomDensv1}
\rh \left(\frac{\p {\bf V}}{\p t} + \left({\bf V}\cdot \nabla \right) {\bf V}\right) &=& - \nabla p + {\bf J} \times {\bf B} \nonumber \\
&&\,- \frac{m_e}{e^2} \left({\bf J}\cdot \nabla\right)\left(\frac{{\bf J}}{n}\right), 
\end{eqnarray}
and the Ohm's law
\begin{eqnarray} \label{Ohmv1}
&&{\bf E} + {\bf V} \times {\bf B} - \frac{{\bf J} \times {\bf B} - \nabla p_e + \delta \nabla p_i}{en} \nonumber \\
&&= \frac{m_e}{ne^2} \left[\frac{\p {\bf J}}{\p t} + \nabla \cdot \left({\bf V} {\bf J} + {\bf J} {\bf V} - \frac{1}{en} {\bf J} {\bf J} \right) \right].
\end{eqnarray}
The variables $\rh$, ${\bf V}$ and ${\bf J}$ serve as the total mass density, centre-of-mass velocity and the current respectively. The variable $n$ appearing in (\ref{MomDensv1}) and (\ref{Ohmv1}) is the number density, and is defined as $n = \rho/\left(m_i + m_e\right)$, with $m_i$ and $m_e$ representing the ion and electron masses respectively. In the above expressions, observe that $\mu_0 {\bf J} = \nabla \times {\bf B}$, the total pressure is represented by $p$ whilst $p_i$ and $p_e$ denote the ion and electron pressures respectively, $\delta = m_e/m_i$ is the mass ratio, and $m_\lambda$ represents the mass of the species `$\lambda$'. Broadly speaking, the above set of equations are derived from the standard two-fluid theory of plasma physics \cite{GP04} by neglecting the displacement current, imposing quasineutrality, and carrying out a systematic expansion in $\delta$. We refer the reader to \cite{L59,GP04,KLMWW14}, where a detailed, and rigorous, derivation of extended MHD from two-fluid theory is presented (see also \cite{AKY15}). The regimes of validity for extended MHD, and the specific conditions under which certain terms can be eliminated to obtain simpler models, are described in \cite{GP04,KM14,KLMWW14}.

If one adopts the standard Alfv\'en units, and introduces the dynamical variable 
\begin{equation}
{\bf B}^\star = {\bf B} + d_e^2 \nabla \times \left[\frac{\nabla \times {\bf B}}{\rh} \right],
\end{equation}
which is well-known from electron MHD \cite{GKR94} and collisionless (two-fluid based) reconnection studies \cite{OP93,CGPPS98}, we observe that (\ref{MomDensv1}) and (\ref{Ohmv1}) can be recast into
\begin{eqnarray}  \label{MomDensv2}
\frac{\p {\bf V}}{\p t} + \left(\nabla \times {\bf V}\right) \times {\bf V} &=& -\nabla \left(h + \frac{V^2}{2}\right) + \frac{\left(\nabla \times{\bf B}\right) \times {\bf B}^\star}{\rh} \nonumber \\
&&\,- d_e^2 \nabla \left[\frac{\left(\nabla \times {\bf B}\right)^2}{2\rh^2}\right],
\end{eqnarray}
\begin{eqnarray} \label{Ohmv2}
\frac{\p {\bf B}^\star}{\p t} &=& \nabla \times \left({\bf V} \times {\bf B}^\star \right) - d_i \nabla \times \left(\frac{\left(\nabla \times{\bf B}\right) \times {\bf B}^\star}{\rh}\right) \nonumber \\
&&\,+d_e^2 \nabla \times \left[\frac{\left(\nabla \times {\bf B}\right) \times \left(\nabla \times {\bf V}\right) }{\rh}\right],
\end{eqnarray}
where the assumption of a barotropic equation of state was used in simplifying the equations. The total enthalpy $h$, in this scenario, is related to the pressure $p$ via the relation $\nabla h = \rho^{-1} \nabla p$, whilst $d_i = c/\left(\omega_{pi} L\right)$ and $d_e = c/\left(\omega_{pe} L\right)$ serve as the normalized electron and ion skin depths respectively. The quantities $\omega_{pi}$ and $\omega_{pe}$ are the ion and electron plasma frequencies respectively, defined via $\omega_{p \lambda} = \sqrt{n_{\lambda 0} q_\lambda^2/\varepsilon_0 m_\lambda}$ with `$\lambda$' denoting the species label. Here, $q_\lambda$ and $m_\lambda$ are the charge and mass of the given species, whilst $n_{\lambda 0}$ is a characteristic number density; for this reason, one must view $d_i$ and $d_e$ as normalization \emph{constants} expressed in terms of the fiducial values of the ion and electron plasma frequencies respectively. The intermediate steps involved in deriving (\ref{MomDensv2}) and (\ref{Ohmv2}) from (\ref{MomDensv1}) and (\ref{Ohmv1}) have been presented in \cite{AKY15}.

Furthermore, it can be shown that (\ref{MomDensv2}) and (\ref{Ohmv2}), in conjunction with (\ref{ContEq}), conserve the energy:
\begin{equation} \label{HamExtMHD}
H = \int_D d^3x\,\left[\frac{\rh V^2}{2} + \rh U(\rh) + \frac{B^2}{2} + d_e^2 \frac{\left(\nabla \times {\bf B}\right)^2}{2\rh} \right].
\end{equation}
Observe that the above expression does depend on $d_e$ but is independent of $d_i$. We observe that the last term in the above expression, proportional to $d_e^2$, is absolutely necessary for energy conservation and emerges via the last term on the RHS of (\ref{MomDensv1}). The latter is often neglected in textbook treatments, leading to erroneous conclusions; see \cite{KM14} for a detailed discussion of the same.

\subsection{Common Hamiltonian structure of the extended MHD models and associated properties} \label{SSecComHam}

We are now in a position to commence our analysis of the different extended MHD models.\\

{\bf Hall MHD:} Hall MHD (HMHD) is a model that neglects electron inertia, and it amounts to letting $d_e \rightarrow 0$ in (\ref{MomDensv2}), (\ref{Ohmv2}) and (\ref{HamExtMHD}). Alternatively, it can be viewed as the model wherein the last term on the RHS of (\ref{MomDensv1}) is neglected, along with the last term in the first line of (\ref{Ohmv1}), and all the terms on the second line of (\ref{Ohmv1}). The current formulation of barotropic Hall MHD was presented in \cite{YH13}, and we shall reproduce it below, as it constitutes a core part of our investigations:
\begin{eqnarray} \label{HallMHDNCPB}
\{F,G\}^{HMHD} &=& - \int_D d^3x\,\Bigg\{\left[F_\rh \nabla \cdot G_{\bf V} + F_{\bf V} \cdot \nabla G_\rh \right] \nonumber \\
&& \quad  - \frac{\left(\nabla \times {\bf V}\right)}{\rh} \cdot \left(F_{\bf V} \times G_{\bf V}\right) \nonumber \\
&& \quad  - \frac{{\bf B}}{\rh} \cdot \left(F_{\bf V} \times \left(\nabla \times G_{\bf B}\right) \right)  \\
&&\quad  + \frac{{\bf B}}{\rh} \cdot \left(G_{\bf V} \times \left(\nabla \times F_{\bf B}\right)\right) \Bigg\} \nonumber \\
&&-\, d_i \int_D d^3x\,\frac{{\bf B}}{\rh} \cdot \left[\left(\nabla \times F_{\bf B}\right) \times \left(\nabla \times G_{\bf B}\right)\right], \nonumber
\end{eqnarray}
and we shall represent this noncanonical bracket as
\begin{equation} \label{HMHDOldVar}
\{F,G\}^{HMHD} = \{F,G\}^{MHD} + \{F,G\}^{Hall},
\end{equation}
where (i) $\{F,G\}^{MHD}$ comprises of the first four lines on the RHS of (\ref{HallMHDNCPB}) and is the ideal MHD bracket first derived in \cite{MG80}, and (ii) $\{F,G\}^{Hall}$ is the last term on the RHS of (\ref{HallMHDNCPB}), and is characterized by the presence of the factor $d_i$. We observe that the Jacobi identity for this (noncanonical) bracket was first shown in \cite{AKY15}, and an alternative, more detailed, version was presented in \cite{LMM15}. Through the suitable use of (\ref{HallMHDNCPB}), it is easy to establish that there are two helicities for Hall MHD, which serve as Casimir invariants of the model; Casimirs are special invariants which satisfy the property $\{F,C\} = 0$ for all arbitrary choices of $F$. The two Casimirs of interest are
\begin{equation} \label{MagHel}
\calc_I = \int_D d^3x\,{\bf A}\cdot{\bf B},
\end{equation}
\begin{equation} \label{CanHel}
\calc_{II} = \int_D d^3x\,\left({\bf A} + d_i {\bf V}\right) \cdot \left({\bf B} + d_i \nabla \times {\bf V}\right),
\end{equation}
which represent the magnetic and ion canonical helicities respectively \cite{Turn86}. Hall MHD exhibits two frozen-in quantities, which behave as Lie-dragged 2-forms. These correspond to dynamical equations of the form
\begin{equation}
\left[\frac{\p}{\p t} + \mathfrak{L}_{\bf V}\right]\left({\bf B} + d_i \nabla \times {\bf V}\right)\cdot d{\bf S} = 0,
\end{equation}
\begin{equation}
\left[\frac{\p}{\p t} + \mathfrak{L}_{{\bf V}_e}\right]{\bf B}\cdot d{\bf S} = 0,
\end{equation}
where $\mathfrak{L}_{\bf X}$ indicates the Lie-derivative with ${\bf X}$ serving as the flow field, ${\bf V}_e = {\bf V} - d_i \nabla \times {\bf B}/\rh$ denotes the electron velocity, whilst ${\bf B}\cdot d{\bf S}$ and $\left({\bf B} + d_i \nabla \times {\bf V}\right)\cdot d{\bf S}$ constitute the magnetic and ion canonical vorticity fluxes respectively; here $d{\bf S}$ represents an area element. For a discussion of the Lie derivative (and the phenomenon of Lie dragging) in ideal HD and MHD, we refer the reader to \cite{TY93,AK98}. The two expressions are equivalent to the statement that the canonical vorticity (curl of the canonical momenta) flux of each species is Lie-dragged by the corresponding velocity of that species. We shall return to this issue in greater detail in Section \ref{SecGTXMHD}. \\

{\bf Inertial MHD:} Inertial MHD (IMHD) arises upon setting $d_i \rightarrow 0$ in (\ref{Ohmv2}). The astute reader may wonder why $d_i \rightarrow 0$ does not automatically imply $d_e \rightarrow 0$ as well since $m_e \ll m_i$. However, we emphasize that the two parameters are \emph{independent}. In particular, inertial MHD is valid when the time scale for changes in the current is much shorter than the electron gyro period \cite{KM14}. A practical use of inertial MHD stems from the fact that a simplified and reduced version yields the famous Ottaviani-Porcelli reconnection model \cite{OP93}. Alternatively, inertial MHD amounts to dropping the terms on the RHS of the first line of (\ref{Ohmv1}). A Hamiltonian formulation for the 2D version was presented in \cite{LMT14}, and the full structure was determined in \cite{AKY15}. An independent bracket for the model was constructed in the latter, but \cite{LMM15} showed that the inertial MHD bracket could be mapped to the Hall bracket as follows:
\begin{equation} \label{HallInertEqv}
\{F,G\}^{IMHD} \equiv \{F,G\}^{HMHD} \left[\mp 2d_e;\,\boldsymbol{\calb}_{\pm}\right],
\end{equation}
and this indicates that the inertial MHD bracket is exactly identical to the Hall MHD bracket provided that ${\bf B} \rightarrow \boldsymbol{\calb}_{\pm} = {\bf B}^\star \pm d_e \nabla \times {\bf V}$ and $d_i \rightarrow \mp 2d_e$ in (\ref{HallMHDNCPB}).  It is evident from (\ref{HallInertEqv}) that there exist \emph{two} different transformations that map the Hall MHD bracket to inertial MHD. 

We see that the new variables $\boldsymbol{\calb}_{\pm}$, which empower us to transition between the two brackets, are closely related to the two helicities of inertial MHD, which have the form
\begin{equation} \label{IMHDHelI}
\calc_{I,II} = \int_D d^3x\, \left({\bf A}^\star \pm d_e {\bf V}\right) \cdot \left({\bf B}^\star \pm d_e \nabla \times {\bf V}\right).
\end{equation}
The difference of the above two helicities leads to the Casimir:
\begin{equation} \label{IMHDHelIII}
\calc_{III} = \int_D d^3x\, {\bf V}\cdot {\bf B}^\star,
\end{equation}
which resembles the cross-helicity invariant of ideal MHD, after performing the transformation ${\bf B} \rightarrow {\bf B}^\star$. Hence, this highlights the commonality of inertial and ideal MHD. Just as in Hall MHD, there exist two Lie-dragged 2-forms, given by
\begin{equation}
\left[\frac{\p}{\p t} + \mathfrak{L}_{{\bf V}_{\pm}}\right]\boldsymbol{\calb}_{\pm}\cdot d{\bf S} = 0,
\end{equation}
where ${\bf V}_{\pm} = {\bf V} \pm d_e \nabla \times {\bf B}/\rh$ and $\boldsymbol{\calb}_{\pm}$ has already been previously introduced. \\

{\bf Extended MHD:} Finally, we consider extended MHD (XMHD), which was first derived correctly in \cite{L59}; the reader is referred to \cite{KM14} where several sources that use incorrect versions of this model are discussed. Extended MHD comprises of (\ref{ContEq}), (\ref{MomDensv2}) and (\ref{Ohmv2}) in their entirety, and its Hamiltonian formulation was first presented in \cite{AKY15}. However, the bracket for extended MHD can be mapped to Hall MHD, as in inertial MHD, as follows:
\begin{equation} \label{ExHallMHDEqv}
\{F,G\}^{XMHD} \equiv \{F,G\}^{HMHD}\left[d_i-2\kappa_{\pm};\,\boldsymbol{\calb}_\pm \right], 
\end{equation}
indicating that the extended MHD bracket is simply recovered via ${\bf B} \rightarrow \boldsymbol{\calb}_\pm:= {\bf B}^\star + \kappa_{\pm} \nabla \times {\bf V}$ and $d_i \rightarrow d_i-2\kappa_\pm$ in (\ref{HallMHDNCPB}), the Hall MHD bracket \cite{LMM15}. We observe that $\kappa_\pm$ is determined via the quadratic equation $\kappa^2 - d_i \kappa - d_e^2 = 0$, implying the existence of two such solutions ($\kappa_{+}$ and $\kappa_{-}$). As a result, it is worth emphasizing that there are \emph{two} possible mappings from the Hall MHD bracket to extended MHD in (\ref{ExHallMHDEqv}). Upon taking the limits $d_i \rightarrow 0$ and $d_e \rightarrow 0$, and mapping to the original variables, it is straightforward to verify that one recovers the inertial and Hall MHD brackets respectively. We also note that the above definition of $\boldsymbol{\calb}_\pm$ reduces to the inertial MHD definition for $\boldsymbol{\calb}_\pm$ when $d_i \rightarrow 0$.

Extended MHD is also endowed with two helicities, given by
\begin{equation} \label{ExtMHDHelIandII}
\calc_{I,II} = \int_D d^3x\, \left({\bf A}^\star + \kappa_\pm {\bf V}\right) \cdot \left({\bf B}^\star + \kappa_\pm \nabla \times {\bf V}\right),
\end{equation}
and one can verify the existence of two Lie-dragged 2-forms, which are governed via
\begin{equation} \label{LieDragXMHD2form}
\left[\frac{\p}{\p t} + \mathfrak{L}_{{\bf V}_{\pm}}\right]\boldsymbol{\calb}_{\pm}\cdot d{\bf S} = 0,
\end{equation}
where ${\bf V}_\pm = {\bf V} - \kappa_\mp \nabla \times {\bf B}/\rh$ and $\boldsymbol{\calb}_\pm = {\bf B}^\star + \kappa_\pm \nabla \times {\bf V}$. We note that ${\bf V}_\pm$ and $\boldsymbol{\calb}_\pm$ duly reduce to their Hall and inertial MHD counterparts upon taking $d_e \rightarrow 0$ and $d_i \rightarrow 0$ respectively.\\

From the preceding analysis, it is possible to draw the following conclusions:
\begin{itemize}
\item There exists a clear hierarchy of models starting from extended MHD. Upon neglecting the Hall terms via $d_i \rightarrow 0$, we arrive at inertial MHD. Similarly, neglecting electron inertia via $d_e \rightarrow 0$ leads to Hall MHD, and neglecting both of them concurrently yields ideal MHD. 
\item This hierarchy is best encapsulated by (\ref{ExHallMHDEqv}) which demonstrates the application of the above limits leads to the emergence of inertial, Hall and ideal MHD brackets from the overarching extended MHD noncanonical bracket.
\item The commonality between all the extended MHD models has been highlighted through the existence of a common bracket, whose basic structure takes the form of (\ref{HallMHDNCPB}). 
\item Owing to this commonality, all extended MHD models are endowed with two helicities, which also serve as Casimir invariants. This feature is clearly inherited from the parent 2-fluid model, whose bracket exhibits similar properties \cite{SK82}.
\item All of the extended MHD models possess two Lie-dragged 2-forms, indicating that generalizations of the frozen-flux condition of ideal MHD can be easily built. Consequently, this implies that these quantities serve as the analogs of the magnetic field in ideal MHD, enabling the generalizations of the Cauchy formula for the latter. 
\item The above property makes it possible to construct a unified (Lagrangian variable) action principle for these models, by building the constraints into the model \emph{a priori}, akin to the ideal MHD action \cite{Newcomb62}. By employing the reduction procedure, we can also derive the noncanonical bracket (\ref{HallMHDNCPB}) in a rigorous manner. We note that both these aspects have been successfully tackled in \cite{DML15}.
\item The unified action principle delineated in \cite{DML15} is complementary to the approach espoused in \cite{KLMWW14}, where an alternative (Eulerian-Lagrangian variable) action principle was studied. 
\end{itemize}

\section{Geometric and topological properties of the extended MHD models} \label{SecGTXMHD}

Hitherto, our analyses have not considered the explicit consequences of the commonalities described in Section \ref{SecHam}, and have, instead, focused primarily on highlighting them. Now, we shall present a few applications of our (unified) Hamiltonian formulation, and highlight its advantages. Henceforth, we shall adopt a coordinate independent language wherever possible, as it simplifies and generalizes our discussion.

\subsection{Generalized circulation and helicity conservation theorems} \label{SSecGenCH}
Firstly, we begin by noting that one can define a generalized vector potential from $\boldsymbol{\calb}_\pm:= {\bf B}^\star + \kappa_{\pm} \nabla \times {\bf V}$ via $\boldsymbol{\calb}_\pm = \nabla \times \boldsymbol{\cala}_\pm:=\nabla \times \left({\bf A}^\star + \kappa_{\pm} {\bf V}\right)$. After some extensive algebra, it is possible to show that
\begin{equation}\label{ApmComp}
	\frac{\partial \boldsymbol{\cala}_\pm}{\partial t} =
	\nabla \boldsymbol{\cala}_\pm \cdot\mathbf{V}_\pm-\mathbf{V}_\pm\cdot\nabla \boldsymbol{\cala}_\pm
	+\nabla\psi_\pm,
\end{equation}
where ${\bf V}_\pm = {\bf V} - \kappa_\mp \nabla \times {\bf B}/\rh$ was defined earlier, and
\begin{equation} \label{Psidefn}
	\psi_\pm:=\kappa_\mp h_e-\Big(\kappa_\pm+\frac{d_e^2}{d_i}\Big) h_i - \phi + \kappa_\mp
	d_e^2\, \frac{J^2}{2\rho}-d_e^2\frac{\mathbf{J}\cdot\mathbf{V}}{\rho}.
\end{equation}
In (\ref{Psidefn}), note that $h_\lambda$ is the enthalpy of species $\lambda$, $\phi$ is the electrostatic potential and ${\bf J}$ is the current. It is more intuitive to rewrite (\ref{ApmComp}) as
\begin{equation}\label{LiedragA}
	\left[\frac{\p}{\p t} + \mathfrak{L}_{{\bf V}_{\pm}}\right]  \mathsf{A}_\pm = d\psi_\pm,
\end{equation}
where $ \mathsf{A}_\pm$ is the 1-form associated with the components of $\boldsymbol{\cala}_\pm$. Similarly, we can introduce the 2-form $ \mathsf{B}_\pm = d  \mathsf{A}_\pm$, whose evolution is determined by applying the exterior derivative `$d$' to (\ref{LiedragA}). We use the the fact that $d^2 = 0$, along with the commutative property of the exterior derivative and the Lie derivative \cite{Arn78}, thereby leading us to the relations
\begin{equation}\label{LiedragB}
	\left[\frac{\p}{\p t} + \mathfrak{L}_{{\bf V}_{\pm}}\right]  \mathsf{B}_\pm = 0,
\end{equation}
and this is identical to (\ref{LieDragXMHD2form}). In other words, in our (new) notation, $ \mathsf{B}_\pm \equiv \boldsymbol{\calb}_{\pm}\cdot d{\bf S}$. Hence, it is possible to undertake a consistency check, and verify that (\ref{LiedragB}) leads to
\begin{equation}
	\frac{\partial \boldsymbol{\calb}_\pm}{\partial t} =
	\nabla\times(\mathbf{V}_\pm \times \boldsymbol{\calb}_\pm),
\end{equation}
upon using $\nabla \cdot \boldsymbol{\calb}_\pm = 0$ and noting that the vector density $\boldsymbol{\calb}_\pm$ is dual to the 2-form $B_\pm$ \cite{TY93}. We can also introduce the 3-form $\calk_\pm =  \mathsf{A}_\pm \wedge d \mathsf{A}_\pm$, which we shall return to shortly hereafter.

From fluid mechanics, the conservation of circulation has been known since the 19th century. It is now straightforward to show that one can derive a generalized circulation theorem.
\begin{eqnarray}
	&& \frac{d}{dt}\int_{L_\pm(t)} \boldsymbol{\cala}_\pm\cdot d\mathbf{l} \,\bigg|_{t=t_0} \nonumber 
	\\
	&& =
	\frac{d}{dt}\int_{L_\pm(t)}  \mathsf{A}_\pm (t) \,\bigg|_{t=t_0} =\frac{d}{dt}\int_{L_\pm(t_0)} \Phi_{{\bf V}_{\pm},t}^* \mathsf{A}_\pm (t)
	\,\bigg|_{t=t_0} \nonumber
	\\
	&& =\frac{d}{dt}\int_{L_\pm(t_0)}  \mathsf{A}_\pm(t) +
	(t-t_0)\mathfrak{L}_{{\bf V}_\pm} \mathsf{A}_\pm + \mathcal{O}\big((t-t_0)^2\big)
	\,\bigg|_{t=t_0}  \nonumber
	\\
	&& =\int_{L_\pm(t_0)}
	\frac{\partial  \mathsf{A}_\pm}{\partial t} +
	\mathfrak{L}_{{\bf V}_{\pm}}  \mathsf{A}_\pm \,\bigg|_{t=t_0}  =\int_{L_\pm(t_0)}
	d\psi_\pm =0, 
\end{eqnarray}
where $\Phi_{{{\bf V}_\pm},t}^*$ denotes the pullback with vector field $\mathbf{V}_\pm$ parametrized by $t$ \cite{AM78}. The integration is carried over the contour $L_\pm(t)$, and the above statement indicates that the generalized vorticity flux is frozen-in for a fluid moving with velocity $\mathbf{V}_\pm$ -- a generalization of the famous frozen-flux condition of ideal MHD. This can be explicitly worked out, as shown below
\begin{eqnarray}
	&& \frac{d}{dt}\int_{S_\pm(t)} \boldsymbol{\calb}_\pm\cdot d\mathbf{S} \,\bigg|_{t=t_0} = \frac{d}{dt}\int_{S_\pm(t)}  \mathsf{B}_\pm (t) \,\bigg|_{t=t_0} \nonumber \\
	&& =
	\int_{S_\pm(t_0)} \frac{\partial  \mathsf{B}_\pm}{\partial t} +
	\mathfrak{L}_{\mathbf{V}_{\pm}}  \mathsf{B}_\pm \,\bigg|_{t=t_0} = 0.
\end{eqnarray}
The 3-forms associated with extended MHD were defined earlier via $\mathcal{K}_\pm:=  \mathsf{A}_\pm \wedge d \mathsf{A}_\pm$, and we emphasize that $K_\pm := \int_{V_\pm} \mathrm{Tr}\left(\calk_\pm\right)$ represent the generalized helicities of ideal MHD, and $\mathrm{Tr}$ denotes the (ad-invariant) inner product. We shall drop this notation $\left(\mathrm{Tr}\right)$ henceforth, but it is implicitly present whenever we deal with helicity-like quantities. We find that (\ref{LiedragA}) can be duly manipulated to yield
\begin{equation}
	\frac{\partial\mathcal{K}_\pm}{\partial t} + \mathfrak{L}_{\mathbf{V}_{\pm}}
	\mathcal{K}_\pm = d\psi_\pm\wedge d  \mathsf{A}_\pm = d(\psi_\pm d  \mathsf{A}_\pm),
\end{equation}
and by invoking Stokes' theorem, we end up with
\begin{equation} \label{HelConsDiff}
	\frac{d}{dt}\int_{V_\pm(t)}\mathcal{K}_\pm = \int_{V_\pm(t)} d(\psi_\pm
	d  \mathsf{A}_\pm) = \int_{\partial V_\pm(t)} \psi_\pm d \mathsf{A}_\pm = 0,
\end{equation}
as long as the generalized vorticity vanishes on the boundary. It is evident that (\ref{HelConsDiff}) constitutes another proof for helicity conservation, thereby complementing the earlier (coordinate dependent) results presented in \cite{AKY15,LMM15,DML15}. It was shown in \cite{PM96,PAM96} - see also \cite{Cal63,Yah95} for associated treatments - that magnetic or fluid helicity conservation was a natural consequence of Noether's theorem on account of the (Lagrangian) particle relabelling symmetry of the ideal HD and MHD actions. By applying a similar procedure to the extended MHD action \cite{DML15}, the invariance of the helicities of extended MHD can be established accordingly.

\subsection{Topological aspects of the generalized helicities of extended MHD} \label{SSecHelXMHD}
Now, we shall take a greater look at the topological ramifications of $K_\pm$ and (\ref{HelConsDiff}), viz. the generalized helicities and their conservation properties respectively.

Let us begin by recalling that $ \mathsf{A}_{+}$ and $ \mathsf{A}_{-}$ serve as 1-forms, appropriately constructed from $\boldsymbol{\cala}_\pm$, where the latter was defined towards the beginning of Section \ref{SSecGenCH}. If one lets $d_e \rightarrow 0$ and $d_i \rightarrow 0$, we have already indicated that the vector potential ${\bf A}$ follows from $\boldsymbol{\cala}_\pm$. Yet, it is important to recognize that all other versions of extended MHD have, not one, but \emph{two} such 1-forms. It is well known that the general expression for a helicity-type quantity is given by $H = \int_\calm P \wedge dP$, where $\calm$ is a compact 3-manifold and $P$ is a 1-form. We have dropped the inner product operator $\left(\mathrm{Tr}\right)$ as noted earlier. Hence, one can duly construct two helicity-like quantities by setting $P =  \mathsf{A}_{\pm}$ and the corresponding (generalized) helicities are given by $K_{\pm}$.

We have reiterated the above steps because the crucial aspect of our work is that these generalized 1-forms, 2-forms and helicities can be seen as the exact analogues of the vector potential/velocity, magnetic field/vorticity, and magnetic/fluid helicity respectively. As a result, we are in the remarkable position of exploiting every known topological property of ideal HD or MHD by generalizing it to extended MHD via the variable transformations introduced here, and in \cite{LMM15}.

For instance, consider the description of the fluid helicity in terms of thin vortex filaments, which are represented collectively by an oriented knot (or link) in $\calm$. The expression for the fluid helicity is given by
\begin{equation} \label{HelDecom}
    H = \sum_i \nu_i^2 Lk_i + 2 \sum_{i\,j} \nu_i \nu_j Lk_{ij},
\end{equation}
where $\nu_i$ denotes the vortex circulation, whilst $Lk_i$ and $Lk_{ij}$ are the self-linking and Gauss linking numbers respectively \cite{MR92,RM92}. Moreover, we observe that $Lk_i = Wr_i + Tw_i$, implying that the self-linking number can be decomposed into its writhing and twisting numbers; the latter duo are topologically relevant in their own right \cite{MR92,CDG01,BP06,SE14}. The decomposition of helicity into its various components has also been verified empirically through a series of ingenious experiments \cite{KI13,SKPKI14,KSI14}, and numerical simulations in dynamos \cite{ATB09}. If we replace the vortex filaments, circulation, etc. by the generalized counterparts (corresponding to $\boldsymbol{\calb}_\pm$), we find that the generalized helicities can be decomposed in a manner exactly identical to (\ref{HelDecom}). 

For all its elegance and utility, the linking number is beset by a number of limitations. The foremost amongst them is that it cannot distinguish between certain topological configurations, such as the Whitehead link and the Borromean rings \cite{Kauf13}. The conventional means of distinguishing between such configurations is via the Massey product \cite{Berg90} and its generalizations \cite{HM02}, or other higher-order invariants \cite{RA94,VBH04,Akh05}. As per the correspondence between ideal MHD (or HD) and the different variants of extended MHD established earlier, we may be able to construct the equivalent (higher-order) topological invariants for the latter class of models. It is at this juncture that we introduce the remarkable insight provided by Witten \cite{Wit89} between topological quantum field theory (TQFT) and knot theory. In particular, Witten demonstrated that the Jones polynomial, a staple of knot theory, could be naturally interpreted in terms of the Chern-Simons action of $\left(2+1\right)$ Yang-Mills theory. The Chern-Simons action for a non-Abelian field theory is given by
\begin{equation} \label{CSAction}
    S = \int_\calm \left(P \wedge dP + \frac{2}{3} P \wedge P \wedge P\right),
\end{equation}
up to constant factors. Now, suppose that the underlying gauge group is Abelian, and this choice eliminates the second term on the RHS of the above expression. Consequently, we are led to the striking result that the helicity is an Abelian Chern-Simons action \cite{JNPP04,BMM06}. As a result, one can employ the versatile mathematical formulations of Chern-Simons theory (a 3-dimensional TQFT) \cite{Ati90,BM94,Dun99} in the realm of plasma and fluid models, thereby opening up a potentially rich and diverse line of future research, as these methods are more sophisticated than standard paradigm of computing the linking number(s); for instance, the Jones polynomial is capable of distinguishing between the Whitehead link and the Borromean rings (which have an identical linking number of zero, as previously mentioned). Despite the inherent mathematical richness of the helicity/Chern-Simons correspondence, it hasn't been sufficiently exploited from a knot-theoretic perspective -- the mathematical works by \cite{AK98,LR12,LR15} on the Jones and HOMFLYPT polynomials in HD and MHD constitute the only such examples of this \emph{specific} line of enquiry. Although \cite{LR12,LR15} utilized the formal equivalence between the fluid (or magnetic) helicity and Abelian Chern-Simons theory, there have been prior studies in high energy physics and topological hydrodynamics that were cognizant of this concept (see e.g. \cite{AK98,JNPP04}). It is also straightforward to apply this framework to non-Abelian magnetofluid models, as briefly stated in \cite{BMM06}.

Thus, we are free to import the results of \cite{AK98,LR12,LR15} in the context of the generalized helicities. In particular, following the mathematical reasoning delineated in \cite{LR12}, we are free to compute the Jones polynomial for a given configuration of the generalized helicity (of which there are two in all). The proof relies on the construction of the skein relations by means of the Kauffman bracket polynomial, and then introducing orientation to obtain the skein relations of the corresponding Jones polynomial. Let us interpret the results from the preceding discussion for the (simpler) case of Hall MHD. One of the Jones polynomials would arise from the magnetic helicity, whilst the other arises from the canonical helicity. The difference of these two helicities is the sum of the cross and fluid helicities. Hence, the associated Jones polynomial, arising from this remainder, would encapsulate the topological properties of the fluid and cross helicities. 

Quite intriguingly, the Chern-Simons forms are \emph{odd-dimensional} differential forms \cite{TF11}, implying that the Chern-Simons action (\ref{CSAction}) is meaningful only for odd dimensions, given that it is proportional to the integral of the Chern-Simons form. In turn, owing to its identification with the generalized helicities, the latter acquire this distinct mathematical structure only in \emph{odd} dimensions. \emph{Ipso facto}, this may imply that helicities (magnetic, fluid or generalized) of this form will naturally emerge in non-relativistic (3D) theories, but not, perforce, in the case of relativistic theories, as they are intrinsically four-dimensional in nature. In particular, we note that relativistic MHD possesses a cross helicity akin to its 3D counterpart, but the 4D version of the conventional (3D) magnetic helicity has proven to be elusive from a Hamiltonian perspective \cite{DAMP15}, although it has been derived through other avenues \cite{Mah03,YKY14,Peg15}.

It must be recognized that knot polynomials are \emph{not} the only means of distinguishing between different topological configurations. Thus, one can easily utilize more powerful mathematical formalisms to study ideal and extended MHD, examples of which include Khovanov and Heegaard Floer homologies, and possibly contact topology on account of its relevance in Legendrian knots \cite{ET05,Gei08}. In the theory of contact structures, one deals with a plane field $\xi$ on a manifold $\calm$, which can be locally represented as the kernel of a 1-form $\alpha$ (the contact form). A necessary condition for the plane field to be a contact structure is that $\alpha \wedge d\alpha$ is non-zero. If we identify $\alpha$ with $ \mathsf{A}_\pm$, it is evident that $\calk_\pm:=  \mathsf{A}_\pm \wedge d \mathsf{A}_\pm$ must be non-zero -- as a result, a potential connection between the generalized helicities (constructed from the integrals of $\calk_\pm$) and contact geometry arises. We also note that the relationship between contact topology and hydrodynamics has already been probed in the context of Beltrami fields by \cite{EG00}.

At this stage, we observe that $\calk_\pm = 0$ also leads to several interesting results that arise from the Frobenius theorem; see for e.g. Theorem 2.2.26 (pg. 93) of \cite{AM78}. The condition $\calk_\pm = 0$ is equivalent to the associated plane field $\xi = \mathrm{ker}\,\alpha$ being closed under the Lie bracket. Mathematically, the latter amounts to the following statement: if $v_1$ and $v_2$ are sections of $\xi$, their Lie bracket $\left[v_1,v_2\right]$ must also be a section of $\xi$. If a plane field is closed under the Lie bracket, the Frobenius theorem implies that $\xi$ is foliated (simply covered) by surfaces (tangent to $\xi$) \cite{Nir73}. Given that the Frobenius theorem has important ramifications for integrability, and the evident connections with the generalized helicities via $\calk_\pm$, we shall defer further investigations to future publications.

Apart from the topological properties of helicity, as seen in isolation, one can also probe its relationship with energy. For instance, a classic result by Moffatt \cite{Moff90} established a relation between the minimum magnetic energy $E_{min}$, the flux $\Phi$ and the volume $V$ of a magnetic flux tube as follows:
\begin{equation} \label{EminPhiV}
    E_{min} = m\Phi^2 V^{-1/3},
\end{equation}
where $m$ depends on the specific properties of the knot, and it is a topological invariant; see also \cite{FH91,Berg93,Ric08} for similar results. When dealing with extended MHD, the magnetic component of the energy density must be transformed from $B^2$ to ${\bf B}\cdot{\bf B}^\star$. As a result, it is natural to ask whether one generalize the result (\ref{EminPhiV}) to extended MHD, and we intend to pursue this line of enquiry in our subsequent works.

The applications we have outlined thus far barely scratch the surface. There are many other results from HD and MHD that can be imported to extended MHD involving helicity. For instance, one such example is helicity injection. This phenomenon has been widely studied in the solar context \cite{NZZ03,JC07} as it has important ramifications, but there have been no studies dealing with generalized helicity injection. We shall leave such subjects for later investigations -- it is our present goal to highlight the correspondence with HD/MHD, thereby paving the way for conducting in-depth research in these areas.

\section{Discussion and Conclusion} \label{SecConc}
In this paper, we have emphasized and exploited the inherent mathematical power of the unified Hamiltonian structure of several extended MHD models. This enterprise was rendered possible owing to the work of \cite{AKY15}, and the unified Hamiltonian (and its underlying action principle) structure was established in \cite{LMM15,DML15}. 

Quite evidently, a host of avenues open up for future analyses. The first, and possibly, the most significant is the derivation of reduced extended MHD models that retain the Hamiltonian properties of the parent model. Such models are likely to be of considerable relevance in reconnection studies, thereby furthering the basic approach adopted in \cite{OP93,CGPPS98,TMWG08,HHM15}. For this reason, it is equally important to conduct a detailed examination of their stability via Hamiltonian methods, analogous to the extensive study of ideal MHD by \cite{AMP13}. We also note the possibility of using extended MHD models to study dynamos and jets \cite{LM15}, as well as helicity injection \cite{FA85}, the last of which appears to be a completely unexplored arena. Although these models are endowed with the ion and electron skin depths, the absence of the corresponding Larmor radii is evident. To rectify this limitation, it is feasible to use the gyromap \cite{MLA14,LiMo14} in the extended MHD context, to develop a gyroviscous theory analogous to the one formulated by Braginskii.

From the unified Hamiltonian structure of these models, we demonstrated that they possess a common class of Casimir invariants - the generalized helicities. Motivated by these helicities, we sought the generalizations of the vorticity (or magnetic field), and thereby established the existence of two Lie-dragged 2-forms. Thus, the whole enterprise demonstrated that the topological properties of these models are a \emph{natural} consequence of their Hamiltonian structure. We believe that this is a vital, but rather unrecognized, fact that merits further attention. By constructing these helicities and 2-forms, we derived properties such as the generalization of Kelvin's circulation theorem in a geometric setting. Moreover, we also showed that these helicities can be viewed as Abelian Chern-Simons theories, and that the methodology introduced by Witten, for gaining insights into topological quantum field theory, could be employed here. Consequently, we concluded that the Jones polynomials may be used to characterize different (generalized vorticity) configurations, serving as a more powerful tool than the standard Gauss linking number used to characterize fluid or magnetic helicity. By introducing such topological methods for characterizing helicity, their relevance in the domains of astrophysics and fusion is self-evident. One such application, of paramount importance, is to deploy these topological methods in gaining a better understanding of solar magnetic fields \cite{Long05}.

In summary, we have used the noncanonical Hamiltonian formulation of extended MHD models to arrive at their common mathematical structure, which manifests itself via the existence of generalized helicities and Lie-dragged 2-forms. These helicities, which are topological invariants, can be further studied through a host of techniques, including the Jones polynomial \cite{AK98,LR12}. From a conceptual point-of-view, our results are elegant, as they exemplify the spirit of unification common to most physical theories. On the other hand, we also believe that the results presented herein possess manifold concrete applications, especially since the helicities serve both as important topological invariants, and crucial mediators of relaxation and self-organization, reconnection, turbulence, and magnetic field generation (dynamos) in fusion and astrophysical plasmas.  

\section*{Acknowledgments}
ML was supported by the NSF Grant No. AGS-1338944 and the DOE Grant No. DE-AC02-09CH-11466. PJM was supported by DOE Office of Fusion Energy Sciences, under DE-FG02-04ER-54742. ML wishes to acknowledge Eric d'Avignon, Lee Gunderson, Stuart Hudson, Timothy Magee, Swadesh Mahajan and Taliya Sahihi for their perceptive insights and remarks.


%

\end{document}